\documentclass[twocolumn, resetfootnote]{aastex7}
\usepackage{xspace}

\newcommand\target{GS~1826$-$238\xspace}

\DeclareRobustCommand{\adso}{\bgroup\markoverwith{\textcolor{magenta}{\rule[.5ex]{2pt}{0.4pt}}}\ULon}

\usepackage{multirow}
\usepackage{CJKutf8, here}
\usepackage{amsmath}

\usepackage{color}
\usepackage{amsmath}
\usepackage{natbib}

\usepackage{threeparttable} 
\usepackage{makecell}

\usepackage{appendix}
\usepackage {booktabs}

\shorttitle{Return of the clocked burster GS~1826$-$238}
\shortauthors{Takeda et al.}

\begin{document}

\title{Return of the Clocked Burster: Exceptionally Short Recurrence Time in \target}
% \title{Return of the clocked burster GS~1826$-$238:\\ discovery of an exceptionally short burst recurrence time with NinjaSat}
% \title{Return of the clocked burster GS~1826$-$238:\\ NinjaSat discovery of a historically short burst recurrence time}
% \title{The Return of GS~1826$-$238: A Clocked Burster Rediscovered by NinjaSat}
% \title{Return of the clocked burster GS~1826$-$238: the fourth epoch discovered with NinjaSat}
% \title{The Return of GS~1826$-$238: 4th clocked burst epoch observed with NinjaSat}
% \title{NinjaSat Observations Reveal the Return of the Clocked Burster GS~1826$-$238}

% \correspondingauthor{Tomohis Takeda}
% \email{tomoshi-takeda@hiroshima-u.ac.jp}

\newcommand{\HU}{Graduate School of Advanced Science and Engineering, Hiroshima University, 1-3-1 Kagamiyama, Higashi-Hiroshima, Hiroshima 739-8526, Japan}

\newcommand{\TUS}{Department of Physics, Tokyo University of Science, 1-3 Kagurazaka, Shinjuku, Tokyo 162-8601, Japan}
\newcommand{\RIKENCPR}{RIKEN Cluster for Pioneering Research (CPR), 2-1 Hirosawa, Wako, Saitama 351-0198, Japan}
\newcommand{\RIKENPRI}{RIKEN Pioneering Research Institute (PRI), 2-1 Hirosawa, Wako, Saitama 351-0198, Japan}
\newcommand{\RIKENRAP}{RIKEN Center for Advanced Photonics (RAP), 2-1 Hirosawa, Wako, Saitama 351-0198, Japan}
\newcommand{\RIKENNishina}{RIKEN Nishina Center, 2-1 Hirosawa, Wako, Saitama 351-0198, Japan}
\newcommand{\KyotoU}{Department of Physics, Kyoto University, Kitashirakawa Oiwake, Sakyo, Kyoto 606-8502, Japan}
\newcommand{\ChibaU}{International Center for Hadron Astrophysics, Chiba University, 1-33 Yayoi, Inage, Chiba 263-8522, Japan}
\newcommand{\iTHEMS}{RIKEN Center for Interdisciplinary Theoretical \& Mathematical Sciences (iTHEMS), RIKEN 2-1 Hirosawa, Wako, Saitama 351-0198, Japan}
\newcommand{\NCUE}{Department of Physics, National Changhua University of Education (NCUE), Changhua 50007, Taiwan}
\newcommand{\CNS}{Center for Nuclear Study (CNS), The University of Tokyo, 7-3-1 Hongo, Bunkyo, Tokyo 113-0033, Japan}
\newcommand{\Kogakuin}{Academi Support Center, Kogakuin University, 65-1 Nakano-machi, Hachioji, Tokyo 192-0015, Japan}
\newcommand{\Rikkyo}{Department of Physics, Rikkyo University, 3-34-1 Nishi Ikebukuro, Toshima-ku, Tokyo 171-8501, Japan}

\author[0009-0008-5133-9131]{Tomoshi Takeda} 
\affiliation{\HU}
\email[show]{tomoshi-takeda@hiroshima-u.ac.jp}
% \email[]{}

\author[0000-0002-8801-6263]{Toru Tamagawa}
\affiliation{\RIKENPRI}
\affiliation{\RIKENNishina}
\affiliation{\TUS}
\email{tamagawa@riken.jp}

\author[0000-0003-1244-3100]{Teruaki Enoto}
\affiliation{\KyotoU}
\affiliation{\RIKENRAP}
% \email[]{enoto.teruaki.2w@kyoto-u.ac.jp}
\email{enoto.teruaki.2w@kyoto-u.ac.jp}

\author[0000-0002-0207-9010]{Wataru Iwakiri}
\affiliation{\ChibaU}
\email{iwakiri@hepburn.s.chiba-u.ac.jp}

\author[0000-0001-8726-5762]{Akira Dohi}
\affiliation{\RIKENPRI}
\affiliation{\iTHEMS}
\email[]{}

\author[0000-0002-6337-7943]{Tatehiro Mihara}
\affiliation{\RIKENPRI}
\email{tmihara@riken.jp}

\author[0000-0001-6314-5897]{Hiromitsu Takahashi}
\affiliation{\HU}
\email{hrtk@hiroshima-u.ac.jp}

\author[0000-0001-8551-2002]{Chin-Ping Hu}
\email{cphu0821@gm.ncue.edu.tw}
\affiliation{\NCUE}

\author[0009-0008-3926-363X]{Amira Aoyama} 
\affiliation{\TUS}
\affiliation{\RIKENPRI}
\email{amira.aoyama@riken.jp}

\author{Naoyuki Ota}
\affiliation{\TUS}
\affiliation{\RIKENNishina}
\email{naoyuki.ota@riken.jp}

\author{Satoko Iwata}
\affiliation{\TUS}
\affiliation{\RIKENNishina}
\email{satoko.iwata@riken.jp}

\author{Takuya Takahashi}
\affiliation{\TUS}
\affiliation{\RIKENNishina}
\email{1225545@ed.tus.ac.jp}

\author{Kaede Yamasaki}
\affiliation{\TUS}
\affiliation{\RIKENNishina}
\email{kaede.yamasaki@a.riken.jp}

\author{Takayuki Kita}
\affiliation{\ChibaU}
\email{takayukikita@hepburn.s.chiba-u.ac.jp}

\author{Soma Tsuchiya}
\affiliation{\TUS}
\email{1222075@ed.tus.ac.jp}

\author{Yosuke Nakano}
\affiliation{\TUS}
\email{2222064@ed.tus.ac.jp}

\author{Mayu Ichibakase}
\affiliation{\Rikkyo}
\email{mayu.ichibakase@a.riken.jp}

% \author[0000-0001-8551-2002]{Chin-Ping Hu}
% \affiliation{\NCUE}
% \email[]{}

\author[0000-0002-0842-7856]{Nobuya Nishimura}
\affiliation{\Kogakuin}
\affiliation{\CNS}
\affiliation{\RIKENPRI}
%\affiliation{\NAOJ}
\email{nobuya@cns.s.u-tokyo.ac.jp}

\collaboration{all}{(NinjaSat collaboration)}

\correspondingauthor{Tomohis Takeda}
% \email{tomoshi-takeda@hiroshima-u.ac.jp}

\begin{abstract}
We report the discovery of an exceptionally short burst recurrence time in the well-known clocked burster GS~1826$-$238, observed with the CubeSat X-ray observatory NinjaSat.
In 2025 May, GS 1826$-$238 underwent a soft-to-hard state transition for the first time in 10 years.
On June 23, NinjaSat began monitoring GS~1826$-$238 in the hard state and continued until it returned to a steady soft state.
During this period, we detected 19 X-ray bursts: 14 during the hard state, 4 in the transitional state, and 1 in the soft state.
In the hard state, we identified a new clocked bursting epoch, during which the burst recurrence time remained highly stable and unprecedentedly short among the clocked bursting phases of GS~1826$-$238, with $t_{\rm rec} = 1.603 \pm 0.040$~hr~($1\sigma$ error).
Previous observations showed that the burst recurrence time in GS~1826$-$238 decreased with increasing mass accretion rate, reached its minimum value of $t_{\rm rec} \sim 3$~hr, and then increased again.
The observed 1.6~hr recurrence time is therefore exceptionally short, indicating anomalous ignition conditions.
We propose that this phenomenon reflects fuel accumulation over a smaller fraction of the neutron star surface, resulting in a higher local accretion rate compared to earlier epochs.
This scenario naturally accounts for the exceptionally short recurrence time, together with the observed reductions during bursts in blackbody normalization (proportional to the emitting area) and fluence.
We also discuss possible contributions from residual heat in the neutron star crust or the presence of an additional soft spectral component.

\end{abstract}


%------------------------
\section{Introduction}
\label{sec:Introduction}
%------------------------
% X線バーストのイントロ (inc. クロックバースター)
% クロックバースターの利点、これまでの観測・理論でどんなことがわかってきたか。Qb
% GS 1826-24 のイントロとこれまでの観測結果・理論研究
% NinjaSat/GMC のイントロ
% イントロの結び

Type I X-ray bursts are explosive transients that occur in low-mass X-ray binaries (LMXBs) and are triggered by unstable thermonuclear burning on the surface of an accreting neutron star (NS) (for a review, see \citealt{Galloway&Keek2021}).
Among over 115 known X-ray bursters~\citep{Galloway2020}, few sources exhibit notably regular burst recurrence times, called clocked bursters, which provide ideal laboratories for the theoretical models.
\target, the best-studied example of the clocked bursters, was initially discovered by the Ginga X-ray satellite~\citep{Makino1988}.
Its remarkably regular burst recurrence time and the highly identical burst profiles were later confirmed by BeppoSAX and RXTE~\citep{Ubertini1999,Galloway2004}.
Previous studies have shown that observation–model comparisons yield valuable constraints on the physical properties of the NS and its companion star~(e.g., \citealt{Heger2007ApJ...671L.141H, 2018ApJ...860..147M, 2020PTEP.2020c3E02D,Johnston2020, Dohi2021ApJ...923...64D}).

The X-ray spectra of NS-LMXBs mainly depend on the X-ray luminosities, which are thought to reflect the mass-accretion rate~(e.g., \citealt{Lewin1997xrb..book.....L, Done2007A&ARv..15....1D}).
At high luminosities ($\gtrsim 10^{37}$~erg~s$^{-1}$), the spectra are dominated by a thermal component, and the system is referred to as being in the soft state, whereas at low luminosities ($\lesssim5\times 10^{36}$~erg~s$^{-1}$), they are dominated by a Comptonized component, corresponding to the hard state.
In many NS-LMXBs, including \target, transitions between spectral states associated with changes in X-ray luminosity or mass accretion rate have been observed (e.g., \citealt{Asai2015PASJ...67...92A}).

In the hard spectral state of \target, the burst recurrence time has been observed to gradually decrease from approximately 6 to 3~hr over a span of roughly 10 years, corresponding to an increase in the mass accretion rate onto the NS~\citep{Cornelisse2003A&A...405.1033C, Galloway2004}.
The relation between burst recurrence time, $t_{\rm rec}$, and the bolometric persistent flux, $F_{\rm bol}$, generally follows a power-law function:
\begin{equation}
     t_{\rm rec} = C F_{\rm bol}^{-\eta},
    \label{eq:eta}
\end{equation}
where $\eta = 1.05 \pm 0.02$ is a power-law index~\citep{Galloway2004}\footnote{A small number of other sources were also found to follow the power-law relation, with $\eta$ values around 1 (see \citealt{Takeda2025PASJ...77L..24T} and references therein). 
We note that theoretical studies suggest that $\eta$ is insensitive to the composition of accreted matter for bursts occurring at moderate accretion rates ($\sim$10\% of the Eddington rate) and that are not helium-rich, as in \target~\citep{Galloway2006ApJ...652..559G, Lampe2016}. In contrast, more compact NS models tend to have lower values of $\eta$~\citep{Dohi+24}.} and $C$ is a constant.
However, \citet{Thompson2008ApJ...681..506T} reported that bursts observed with RXTE during the 2003 April epoch occurred approximately 30\% earlier than predicted based on the measured value of $F_{\rm bol}$.
Simultaneous XMM-Newton observations revealed an additional soft X-ray component below 2~keV, contributing 30--50\% of $F_{\rm bol}$. 
This suggests that the deviation was caused by an underestimation of $F_{\rm bol}$ due to RXTE’s limited energy coverage (typically 3--100~keV combined for the PCA and HEXTE), rather than a true change in the amount of burst fuel or ignition conditions.

\target had remained in the hard spectral state since its discovery, except for a brief soft-state transition in 2014 June~\citep{Nakahira2014ATel.6250....1N, Asai2015PASJ...67...92A, Chenevez2016ApJ...818..135C}.
In 2015 July, the source entered a prolonged soft state, during which the burst recurrence time became irregular and the burst profiles deviated significantly from earlier clocked bursting phases~\citep{Ji2018MNRAS.474.1583J,Sanchez2020A&A...634A..58S, Yun2023ApJ...947...81Y, Grefenstette2025ApJ...987..180G}.

In 2025 May, \target underwent a hard-state transition for the first time in 10 years. 
Despite the lack of alerts from all-sky monitor satellites, the CubeSat X-ray observatory NinjaSat~\citep{Enoto2020, Tamagawa2025PASJ...77..466T} began pointing observations of the source on 2025 June 23, utilizing the operational flexibility of a small satellite. 
This monitoring continued until 2025 July 18, by which time the source appeared to have returned to a steady soft state.
We successfully captured this transient hard-state episode and confirmed the return of the periodic burst behavior, with a recurrence time of approximately 1.6~hr, which is the shortest recurrence time observed to date during the clocked phases of \target~\citep{Iwata2025ATel17245....1I}.
Following the NinjaSat report, wide-field monitoring by SVOM/ECLAIRs also confirmed a comparable recurrence time~\citep{2025ATel17251....1B}.

In this Letter, we report the NinjaSat discovery of this unprecedentedly short recurrence time in the clocked burster \target and discuss possible interpretations. 
Unless otherwise noted, all uncertainties are quoted at the $1\sigma$ confidence level throughout this Letter.

%------------------------
% \section{Observation and Data Analysis}
\section{Observation and Data Reduction}
\label{sec:observation}
%------------------------
% NinjaSat 観測の概要・経緯
% GS 1826-24, Crab, BKGD
% イベントカット条件と姿勢補正、レート
% 較正関係 : BKGD, Crab 補正 (レスポンス), ゲイン補正
% MAXI, Swift のデータの出所と Crab 変換

NinjaSat is a 6U-size ($10 \times 20 \times 30$~cm$^3$) CubeSat X-ray observatory.
Since its launch in 2023 November, NinjaSat has successfully demonstrated CubeSat capabilities, enabling the long-term monitoring of bright X-ray sources~\citep{Takeda2025PASJ...77L..24T} and prompt response for X-ray transient events~\citep{Aoyama2025ApJ...986L..29A}.
NinjaSat is equipped with two non-imaging Xe-based Gas Multiplier Counters (GMCs), each covering the 2--50 keV band with an effective area of 16~cm$^2$ at 6~keV~\citep{Tamagawa2025PASJ...77..466T, Takeda2023JINST}.

We used only GMC1 of the two GMCs for the present observations. 
The scientifically available exposure (Good Time Intervals, GTIs) was 206.8~ks from 2025 June 23 (MJD 60849) to July 18 (MJD 60874) for GS 1826-238.
To verify the detector energy calibration and assess the background characteristics during the \target observations, we conducted additional observations of the Crab Nebula and a blank-sky field located at (R.A., Decl.) = ($138\fdg00$, $15\fdg00$), known as BKGD\_RXTE3~\citep{Jahoda2006ApJS..163..401J, Remillard2022AJ....163..130R}. 
These observations were performed during the latter part of the \target monitoring campaign, specifically from MJD 60867 to MJD 60870 for the Crab Nebula, and from MJD 60858 to MJD 60863 for the blank-sky field.
The GTI exposures of the Crab Nebula and the blank-sky region were 29.4 and 59.0~ks, respectively. 
Photon arrival times were corrected to the solar system barycenter with FTOOLS {\tt barycen} using the DE405 planetary ephemeris with source coordinates R.A. $= 277\fdg3675$, Decl.$= -23\fdg7969$~\citep{Barret1995A&A...303..526B}.

Following the method of \citet{Takeda2025PASJ...77L..24T}, we extracted GMC cleaned events for each source.
We corrected the X-ray reduction rate due to the passive collimator of the GMC, which limits the field-of-view (FoV) to $2\fdg1$ at the full-width at half-maximum (FWHM), considering the satellite pointing fluctuations during each observation.

Prior to subsequent X-ray spectral analysis of \target, we verified the GMC detector response by following the procedures outlined in \citet{Aoyama2025ApJ...986L..29A}.
We fitted the Crab Nebula spectrum in the 2--20~keV band using an absorbed power-law model ({\tt tbabs $\times$ powerlaw} in XSPEC), adopting the recent NuSTAR absolute flux measurements as the calibration reference~\citep{Madsen2022JATIS...8c4003M}
\footnote{
Column density $N_{\textrm{H}}=2.2\times 10^{21}$~cm$^{-2}$, photon index $\Gamma = 2.103 \pm 0.001$, its normalization $K = 9.69 \pm 0.02$~photons~cm$^{-2}$~s$^{-1}$~keV$^{-1}$ at 1~keV, and 2--10~keV X-ray flux $F_{2-10} = 2.09 \times 10^{-8}$~erg~s~$^{-1}$~cm$^{-2}$.}.
When fitting the Crab spectrum with the column density fixed to the NuSTAR value ($N_{\textrm{H}}=2.2\times 10^{21}$~cm$^{-2}$), the normalization was found to be approximately 22\% lower than the reference, comparable to the calibration offset reported by \citet{Aoyama2025ApJ...986L..29A}.
We therefore applied a uniform 22\% reduction to the effective area of GMC1 across the entire energy range throughout this study.
After applying the correction, the best-fit parameters obtained with NinjaSat were a photon index of $\Gamma = 2.101 \pm 0.007$, its normalization of $K = 9.65 \pm 0.13$~photons~cm$^{-2}$~s$^{-1}$~keV$^{-1}$ at 1~keV, and the 2--10~keV X-ray flux of $F_{2-10} = (2.093 \pm 0.016) \times 10^{-8}$~erg~s~$^{-1}$~cm$^{-2}$.
These results are consistent with those of NuSTAR within statistical uncertainties, and furthermore, the best-fit values agree to better than 0.4\% for all parameters.

%------------------------
\section{Results}
% \section{Analysis}
% \label{sec:Analysis}
%------------------------

% 長期的なステート遷移。NinjaSat 観測全体 (Fig.1)とステート遷移 CCD。バーストの受かった個数とステートの関係
% バーストライトカーブの解析と特徴づけ, 再帰時間
% 定常放射成分の解析 (ハードステート/ソフトステート)
% バーストの時間分解フィット

\subsection{Spectral State Transition}
\label{sec:state}

\begin{figure*}[ht]
    % \vspace{-15pt}
    \begin{center}
    \includegraphics[width=180mm]{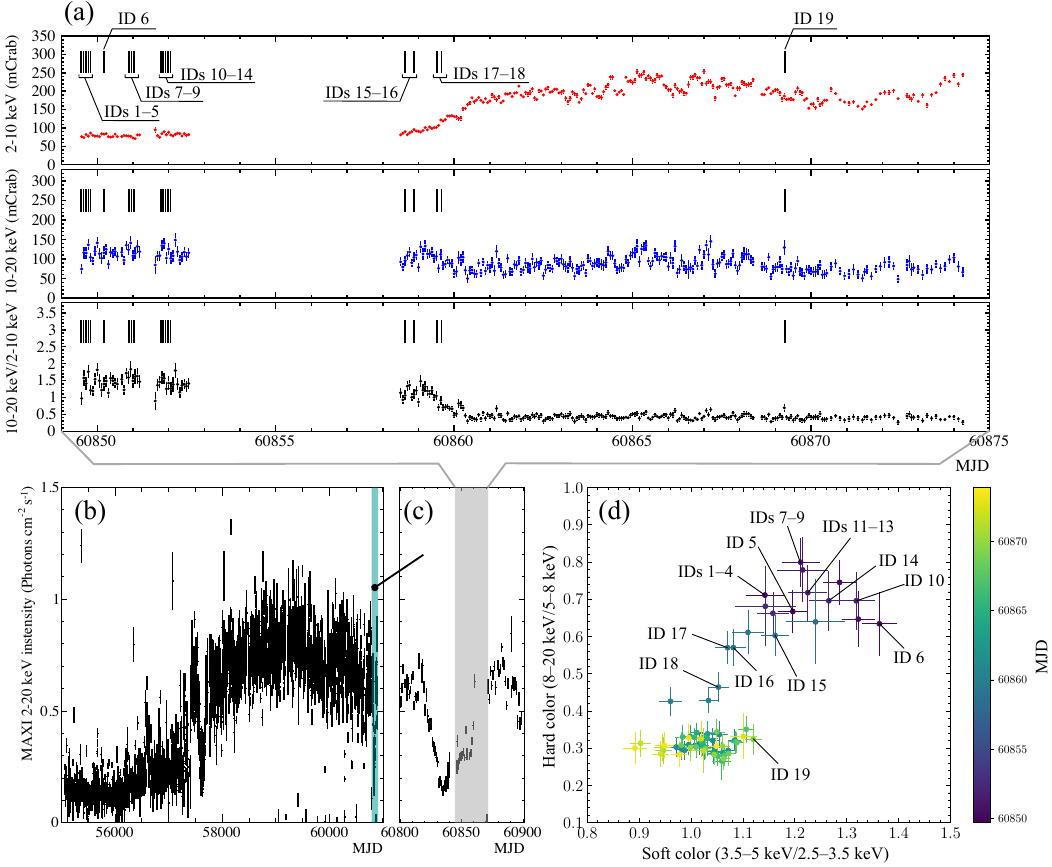}
    \end{center}
    \caption{
    (a)
    NinjaSat light curves of \target in 2--10~keV, 10--20~keV, and their hardness ratio with a bin size of 1.5~hr.
    The time intervals between 50-s before and 300-s after the burst onset are excluded to show the persistent flux, while these burst onsets are indicated by black vertical lines in each panel.
    (b) 
    MAXI/GSC light curve in the 2–20 keV energy range with 1-day bins, downloaded from the MAXI on-demand archive~\citep{Mihara2024}.
    (c) 
    Same as (b), but zoomed in on the latter part. 
    This period is highlighted in panel (b) by the shaded teal region, while the shaded gray region corresponds to the NinjaSat observation interval.
    (d)
    Color-Color diagram observed during the  NinjaSat observation campaign with a bin size of 6.0~hr.
    The IDs of the observed bursts are indicated at corresponding data points.
    }
    \label{fig:light_curves}
\end{figure*}

Figure~\ref{fig:light_curves}(a) shows the 2--10~keV and 10--20~keV light curves of \target observed with NinjaSat, along with the corresponding hardness ratio (HR).  
To place the present observations in the broader context of the spectral state evolution of \target since 2009, we present the long-term monitoring data from the MAXI/GSC~\citep{Matsuoka2009} in Figure~\ref{fig:light_curves}(b, c), highlighting the NinjaSat observation period.
\target had been in the soft state since 2015 and transitioned to the hard state starting around MJD 60820 for the first time in 10~years.
By MJD 60849, when NinjaSat began observations, \target had already completed its transition to the hard state.

To obtain a coarse characterization of changes in the persistent spectrum over our observing period, we constructed a color–color diagram (CCD) from the NinjaSat data, using 6.0~hr bins (Figure~\ref{fig:light_curves}(d)).
The hard and soft colors are defined as the ratios of background-subtracted counts in the (8.0--20.0)/(5.0--8.0)~keV and (3.5--5.0)/(2.0--3.5)~keV bands, respectively.
\target showed a constant flux across all energy bands during the early phase of NinjaSat observations (MJD 60850--60853), during which the HR also remained nearly constant.
Although there was a relatively long observation gap from MJD 60853 to MJD 60858 due to a satellite communication issue with the S-band transceiver, observations resumed successfully after a system reboot.
NinjaSat detected a re-brightening in the 2--10~keV band immediately after the resumption, whereas the 10--20~keV intensity began to gradually decline.
Over the following two days, this contrasting behavior led to a steady decrease in the HR, clearly indicating a spectral state transition.
After this transition, \target remained in the soft state throughout the NinjaSat observation period, as indicated by its consistently low HR and position in the CCD.

\subsection{Type-I X-ray Burst}
\label{sec:burst}

{
\tabcolsep = 3.1pt
\begin{table*}[t]
\centering
% \small
\footnotesize
% \scriptsize
\begin{threeparttable}

\caption{Properties of Type-I X-ray bursts from GS 1826$-$238 observed with NinjaSat.
}
\label{tab:burst_prop}
\begin{tabular*}{\textwidth}{@{\extracolsep{\fill}}ccccccccc}
% \begin{tabular}{cccccccc}
% \begin{tabular}{c*{8}{c}}
\hline
\hline
ID & 
MJD&
$\Delta t_{\rm pre}$ (hr) &
$t_{\rm rise}$ (s) &
$\tau_{\rm D}$ (s) &
$A$ (counts~s$^{-1}$) &
$c_{\rm per}$ (counts~s$^{-1}$) &
Fluence (10$^{-6}$~erg~cm$^{-2}$) &
% \begin{tabular}{c} $\Delta t_{\rm pre}$\\(hr) \end{tabular} & 
% \begin{tabular}{c} $t_{\rm  rise}$\\(s)\end{tabular} & 
% \begin{tabular}{c} $\tau_{\rm D}$ \\(s) \end{tabular}& 
% \begin{tabular}{c} $A$ \\(counts~s$^{-1}$) \end{tabular}& 
% \begin{tabular}{c} Fluence\\(erg cm$^{-2}$) \end{tabular}& 
Note\\
\hline
1 & 60849.55227 &   -   & 5.34 $\pm$ 0.06 & 37 $\pm$ 4 & 8.1 $\pm$ 0.8 & 1.32 $\pm$ 0.09 & $0.827_{-0.06}^{+0.06}$& - \\
2 & 60849.61941 & 1.6114 & 4.0 $\pm$ 1.6 & 29 $\pm$ 3 & 10.2 $\pm$ 1.0 & 1.18 $\pm$ 0.09 & $0.81_{-0.06}^{+0.07}$ &  - \\
3 & 60849.68634 & 1.6065 & 7.1 $\pm$ 2.2 & 36 $\pm$ 5 & 8.1 $\pm$ 0.8 & 1.27 $\pm$ 0.10 & $0.79 _{-0.07}^{+0.07}$&  - \\
4 & 60849.75328 & 1.6064 & 6.7 $\pm$ 1.5 & 52 $\pm$ 8 & 7.6 $\pm$ 0.8 & 1.17 $\pm$ 0.11 & - & Decay \\
5 & 60849.81837 & 1.5622 & 8.5 $\pm$ 2.1 & 34 $\pm$ 5 & 8.4 $\pm$ 0.9 & 1.39 $\pm$ 0.12 & - &  Decay \\
6 & 60850.19907 & 9.1368 & - & - & - & - & - &  Decay \\
7 & 60850.89687 & 16.7471 & 5.1 $\pm$ 1.5 & 52 $\pm$ 5 & 8.2 $\pm$ 0.6 & 1.04 $\pm$ 0.10 & $1.02_{-0.07}^{+0.08}$&  - \\
8 & 60850.96602 & 1.6596 & 8.02 $\pm$ 0.05 & 46 $\pm$ 1 & 6.8 $\pm$ 0.2 & 1.12 $\pm$ 0.03 & $0.68_{-0.05}^{+0.06}$ &  - \\
9 & 60851.03233 & 1.5915 & - & - & - & - & - &  Decay \\
10 & 60851.78662 & 18.1029 & - & - & - & - & - &  Rise \\
11 & 60851.85254 & 1.5821 & 4.45 $\pm$ 0.02 & 38 $\pm$ 5 & 7.5 $\pm$ 0.8 & 1.25 $\pm$ 0.10 & ${0.79}_{-0.06}^{+0.05}$ &  - \\
12 & 60851.91971 & 1.6121 & 4.1 $\pm$ 1.6 & 34 $\pm$ 3 & 9.3 $\pm$ 0.8 & 1.03 $\pm$ 0.09 & ${0.75}_{-0.05}^{+0.06}$ &  -  \\
13 & 60851.98364 & 1.5342 & 8.9 $\pm$ 3.6 & 47 $\pm$ 6 & 7.0 $\pm$ 0.7 & 1.15 $\pm$ 0.10 & ${0.8}_{-0.05}^{+0.05}$ &  -  \\
14 & 60852.05321 & 1.6698 & 6.7 $\pm$ 1.2 & 48 $\pm$ 5 & 9.8 $\pm$ 0.8 & 1.20 $\pm$ 0.11 & ${1.05}_{-0.06}^{+0.07}$ &  -  \\
15 & 60858.63068 & 157.8593 & 6.25 $\pm$ 0.04 & 43 $\pm$ 1 & 7.1 $\pm$ 0.1 & 0.97 $\pm$ 0.02 & - &  Decay  \\
16 & 60858.88376 & 6.0738 & 6.8 $\pm$ 3.3 & 17 $\pm$ 3 & 6.6 $\pm$ 1.0 & 1.21 $\pm$ 0.08 & - &  Decay  \\
17 & 60859.52409 & 15.3680 & 6.5 $\pm$ 1.2 & 22 $\pm$ 3 & 11.7 $\pm$ 1.1 & 1.26 $\pm$ 0.11 & ${0.71}_{-0.05}^{+0.05}$ &  -  \\
18 & 60859.64992 & 3.0199 & 13.4 $\pm$ 2.7 & 11 $\pm$ 2 & 9.5 $\pm$ 1.2 & 1.50 $\pm$ 0.08 & ${0.32}_{-0.04}^{+0.04}$ &  -  \\
19 & 60869.27919 & 231.1026 & 2.8 $\pm$ 0.8 & 9 $\pm$ 2 & 18.5 $\pm$ 3.2 & 2.14 $\pm$ 0.09 & ${0.46}_{-0.05}^{+0.05}$ &  -  \\
\hline
\end{tabular*}
\begin{tablenotes}
\item \textbf{Note.} 
The MJD column lists the burst onset times in Modified Julian Date (MJD).
$\Delta t_{\rm pre}$ represents elapsed time (hr) since the previous burst detected with NinjaSat.
$t_{\rm rise}$ is the time to reach the peak from the onset in a unit of seconds.
Columns $\tau_{D}$, $A$, and $C$ represent the exponential decay timescale, the peak count rate, and the count rate of the constant component, respectively.
Burst fluence in the 0.1--200 keV energy range is estimated from the time-integrated spectral fitting (see text for more detail). 
The Note column indicates whether the “Rise'' or “Decay'' phase of the burst profiles was truncated due to the observational time window.
\end{tablenotes}
\end{threeparttable}
\end{table*}
}

During the NinjaSat observations, 19 events (IDs 1--19) characterized by a rapid rise and exponential decay in X-ray intensity were detected, consistent with the typical morphology of Type-I X-ray bursts.
The first 14 bursts (IDs 1--14) were observed between MJD 60849 and 60852 in the hard state, the last burst (ID 19) occurred in the soft state, and the rest during the transition period from the hard to the soft state, which is illustrated in the light curves in Figure~\ref{fig:light_curves}(a) and the CCD in Figure~\ref{fig:light_curves}(d).

To characterize the burst profile ($f(t)$ as a function of time, $t$), we fitted each light curve with a linear-rise exponential-decay model, described by
{
% \fontsize{8}{13}
\begin{equation}
    \begin{array}{l}
    \leftline{$f(t)$ =} \\
        \begin{cases}
            c_{\rm per} & t \leq t_0 \\
            \frac{A}{t_{\rm rise}}(t-t_0) + c_{\rm per} & t_0 < t \leq t_0 + t_{\rm rise} \\
            A \exp\left(-\frac{t-t_0 - t_{\rm rise}}{\tau_{D}}\right) + c_{\rm per} & t > t_0 + t_{\rm rise}
        \end{cases},
    \end{array}
    \label{eq:profile_burst}
\end{equation}
}where $c_{\rm per}$ is the count rate of the constant component~($\rm counts~s^{-1}$), $A$ is the burst amplitude~($\rm counts~s^{-1}$), $t_0$ is the burst onset time~(s), $t_{\rm rise}$ is the time to reach the peak from the onset~(s), and $\tau_{D}$ is the decay time constant~(s). 
% The corresponding best-fit parameters are given in table \ref{tab:burst_prop}. 
Seven out of the 19 burst profiles were truncated in the rise or decay phase due to the limited observational time window. 
The best-fit parameters for bursts with complete profiles are listed in Table~\ref{tab:burst_prop}. 
For truncated bursts, if the decay phase is covered up to at least twice the exponential decay time, we also include the corresponding parameters in the table. 
Otherwise, only the onset times are provided.
% The best-fit parameters for the bursts with complete profiles are listed in Table~\ref{tab:burst_prop}, whereas only the burst onset times are provided for the truncated ones.

% バーストの説明, Rise, Decay, ステートごとの比較
The bursts in the hard state (IDs 1--14, except 10) have similar profiles.
The average rise time, decay time constant, and peak count rate are $t_{\rm rise} = 6.3 \pm 1.7$~s, $\tau_{\rm D} = 41.1 \pm 7.6$~s, and $A = 8.3 \pm 1.1$ $\rm counts~s^{-1}$, respectively.
On the other hand, the burst in the soft state (ID 19) exhibits a faster rise and decay time of $t_{\rm rise} = 2.8 \pm 0.8$~s and $\tau_{\rm D} = 9 \pm 2$~s, respectively, and a higher peak count rate of $A = 18.5 \pm 3.2$~$\rm counts~s^{-1}$.

Burst onset times are determined with an accuracy of approximately 1~s, and the elapsed times since the previous burst ($\Delta t_{\rm pre}$) range from $1.5342$~hr to $231.1026$~hr. 
During the hard state, the average time interval between bursts (i.e., the burst recurrence time) is 1.603~hr with a standard deviation of 0.040~hr, excluding the significantly longer intervals between IDs 5--6, 6--7, and 9--10.
Note that the burst recurrence time of 1.6~hr is comparable to the orbital period of NinjaSat ($\sim 92$~min), raising the possibility that the apparent recurrence time may be an observational artifact---if bursts occurring at submultiples of 1.6~hr (e.g., 1/2, 1/3, or 1/4 of 1.6~hr) were missed due to observational gaps.  
To test this possibility, we compared the GTIs of NinjaSat with the predicted burst times under such aliasing scenarios and found no consistent alignment, ruling out these scenarios.
Furthermore, SVOM/ECLAIRs observations later confirmed a recurrence time of approximately 1.6~hr \citep{2025ATel17251....1B}, lending further support to our conclusion that the burst recurrence time in the hard state was stable, with $t_{\rm rec} = 1.603 \pm 0.040$~hr, which is the shortest observed during the clocked bursting phases of \target.
After the spectral state transition on MJD~60858, the intervals between bursts became longer than 3~hr, significantly exceeding the 1.6~hr recurrence time observed in the hard state. 
This suggests that the recurrence time also increased, although we cannot fully rule out the possibility of missed bursts between each detected burst, which would result in an apparently longer recurrence time.

\subsection{Persistent Emission}
\label{sec:persistent}

\begin{figure}[t]
    % \vspace{-15pt}
    \begin{center}
        \includegraphics[width=85mm]{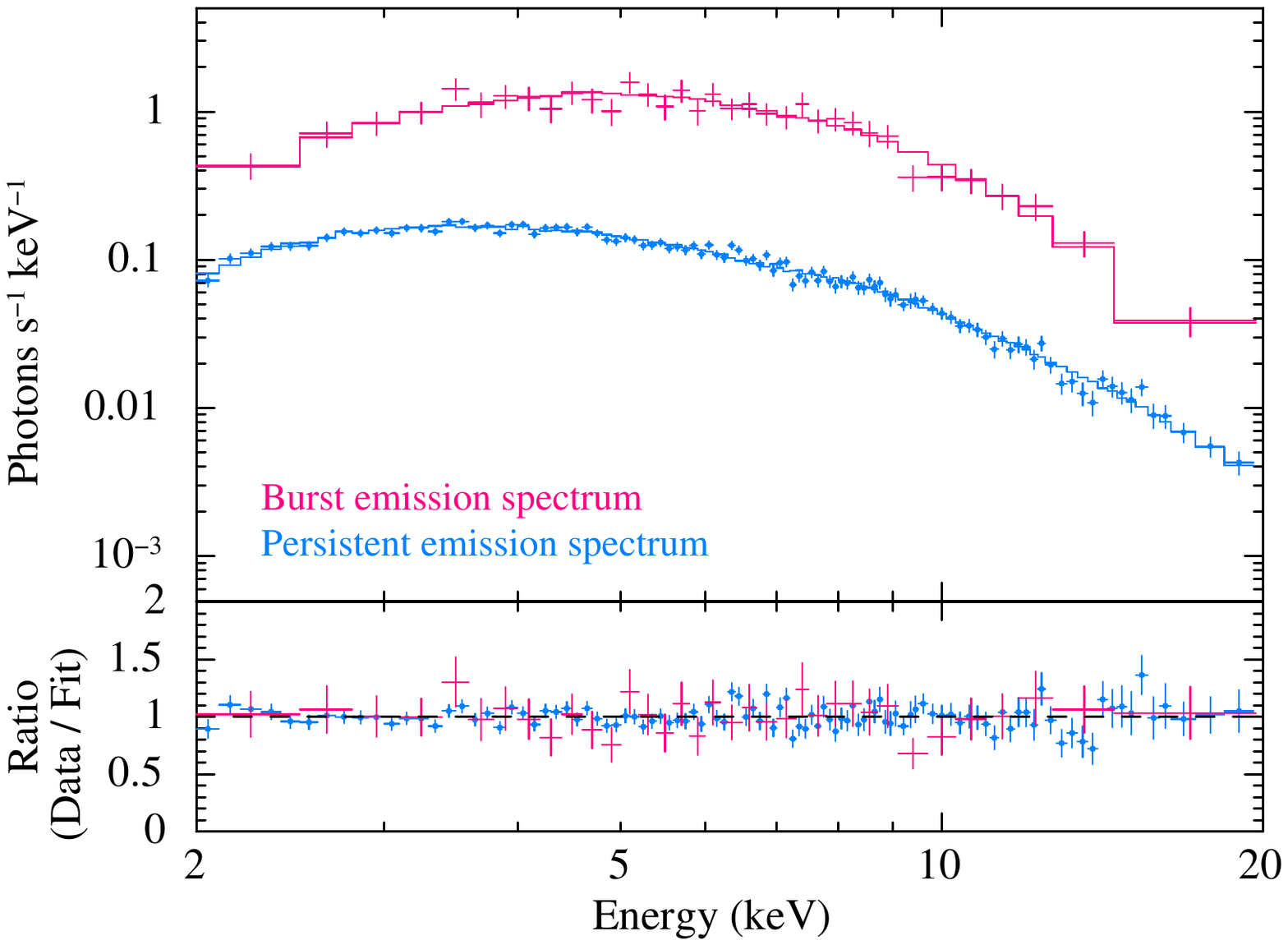}
    \end{center}
    \caption{
    Average energy spectra of the persistent emission (blue circles) and burst emission (pink crosses) in the hard state (IDs 1--14 except 10), along with their respective best-fit models (top), and the corresponding data-to-model ratios (bottom). 
    The burst spectrum was extracted using the interval 10--20~s after burst onset and rebinned to maintain a minimum significance of $5\sigma$ per bin for display purposes.
    % The burst spectrum was extracted using the interval 10--20~s after burst onset.
    % The burst spectrum was extracted from data obtained 10--20~s after burst onset. 
    % The spectral fit results of the persistent and burst emissions (10--20~sec after onset) in the hard state, with their best-fit models (top), and the data-to-model ratios (bottom).
    % The persistent spectrum is shown in pink, while the burst is shown in blue. 
    % The detailed parameters are presented in the text.
    }
    \label{fig:spec}
\end{figure}

% 今回、これ以降のセクションでは hard state でのパーシステントとバーストの解析に focus する。
% We define intervals for spectral extraction from the hard state.
As shown in Section~\ref{sec:state}, the persistent emission during the hard state remained nearly constant.  
To improve the statistics, we integrated the spectra between the onset times of the first (ID~1, MJD 60849.54638) and last (ID~14, MJD 60852.04731) bursts in the hard state, excluding the intervals from 50~s before to 300~s after each burst onset. The total exposure time is 25.9~ks.

The X-ray spectral fit was performed using XSPEC v12.13.1, over the 2--20~keV energy range.  
Following several previous studies (e.g., \citealt{Cocchi2010A&A...509A...2C, Yun2023ApJ...947...81Y}), we fitted the hard-state spectrum using a single Comptonization model, {\tt tbabs $\times$ compTT} in XSPEC.  
The neutral absorption column density was fixed at $0.4 \times 10^{22}$~cm$^{-2}$ \citep{in'tZand1999A&A...347..891I, Pinto2010A&A...521A..79P}, since NinjaSat is not highly sensitive to such low levels of absorption.

Figure~\ref{fig:spec} shows the spectral fit to the persistent emission in the hard state, along with the data-to-model ratio.
A single Comptonization component provides an acceptable fit, with a reduced $\chi^2 = 1.05$ for 102 degrees of freedom. 
The spectrum is characterized by a seed photon temperature constrained to $<0.38$~keV ($0.23_{-0.23}^{+0.15}$~keV) and an electron temperature of $6.1_{-0.9}^{+1.7}$~keV.
The corresponding optical depths are $4.72_{-0.6}^{+0.4}$ for a disk geometry and $10.1_{-1.2}^{+0.9}$ for a spherical geometry.
These parameters lie between those observed in the soft and hard states by NuSTAR \citep{Yun2023ApJ...947...81Y}, consistent with the fact that the source was observed just prior to the state transition.
We then used the {\tt cflux} convolution model in XSPEC to calculate the absorbed 3--25~keV and unabsorbed bolometric 0.1--1000~keV fluxes, which are the energy ranges commonly used in studies of this source (e.g., \citealt{Galloway2017}).
The resulting values are $F_{\rm per} = 2.707_{-0.083}^{+0.056} \times 10^{-9}$~erg~cm$^{-2}$~s$^{-1}$ and $F_{\rm bol} = 4.39_{-0.45}^{+0.44} \times 10^{-9}$~erg~cm$^{-2}$~s$^{-1}$, respectively.

To assess the potential contribution of soft thermal emission below 2~keV~(see Section~\ref{sec:Introduction}), we added a multicolor disk blackbody ({\tt diskbb} in XSPEC) to the original Comptonization model, with its temperature tied to the seed photon temperature of the Comptonization component. 
If such a low-temperature ($\sim$0.2~keV) component were present, the best-fit bolometric flux would increase by $\sim$86\%. 
However, since the Comptonization model already provides an acceptable fit to the spectrum above 2~keV (Figure~\ref{fig:spec}), the additional component is not fully constrained by the data. 
This estimate should therefore be regarded as indicative of the possible contribution of soft emission, rather than a precise measurement. 
The evolution of the persistent emissions after the state transition on MJD 60858 is outside the scope of this study, and we leave it as future work.

\subsection{Time-resolved Burst Spectroscopy}
\label{sec:time-resolved}

To further characterize the burst in the hard state, we performed time-resolved spectroscopy starting from the burst onset to 150~s after the onset. 
The relatively low number of counts observed by NinjaSat prevented time-resolved spectral analysis of individual bursts.
Therefore, we stacked 13 bursts observed in the hard state (IDs 1--14, excluding 10) and generated dynamically binned spectra, each containing at least 200~photons. 
We fixed the parameters of the persistent emission model (Section~\ref{sec:persistent}) and added a thermal blackbody component to account for the burst emission, using the model {\tt tbabs $\times$ (compTT + bbodyrad)} in XSPEC.
The C-statistic \citep{Cash1979ApJ...228..939C} was employed to fit the spectra due to their relatively low photon counts.
As an example, a spectrum accumulated over 10--20~s near the burst peak is shown in Figure~\ref{fig:spec}.

\begin{figure}[t]
    % \vspace{-15pt}
    \begin{center}
        \includegraphics[width=85mm]{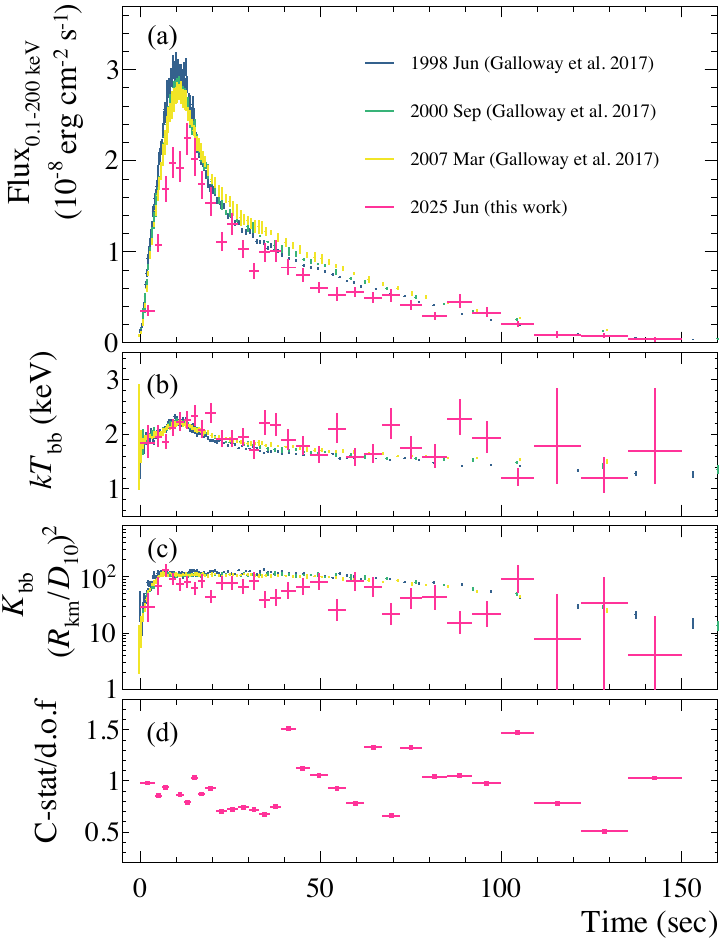}
    \end{center}
    \caption{
    Evolution of the blackbody components during the Type-I X-ray burst from time-resolved spectroscopy, including data from the previous three clocked bursting epochs (\citealt{Galloway2017}\footnote{\url{https://burst.sci.monash.edu/reference}}).
    (a) bolometric (0.1--200~keV) unabsorbed flux in units of 10$^{-8}$~erg~cm$^{-2}$~s$^{-1}$; (b) blackbody temperature, $kT_{\rm bb}$~(keV); 
    (c) blackbody normalization, $K_{\rm bb}$~${(R_{\rm km}/D_{10})^2}$, where $R_{\rm km}$ represents the source radius in kilometers and $D_{10}$ indicates the distance to the source in units of 10 kpc; 
    % (c) blackbody radius, $R_{BB}$~(km) assumming the distance of 6.1~kpc~\citep{Galloway2017}; 
    (d) reduced C-stat value for the fit.
    }
    \label{fig:BBcomponents_vs_time}
\end{figure}

Figure~\ref{fig:BBcomponents_vs_time} shows time-resolved spectroscopic results, alongside those from the previous three clocked bursting epochs reported in \citet{Galloway2017}.
The bolometric unabsorbed 0.1--200~keV flux shown in Figure~\ref{fig:BBcomponents_vs_time}(a) was derived using the convolutional {\tt cflux} model.
The data do not show evidence of photospheric radius expansion (PRE).
The burst reaches a peak flux of $F_{\rm pk} = (2.25 \pm 0.18) \times 10^{-8}$~erg~cm$^{-2}$~s$^{-1}$. 
This value is approximately 20\% lower than measurements from previous epochs, which can be attributed to smaller blackbody normalization.
The burst fluence released up to 150~s is $E_{\rm b} = (0.849 \pm 0.019) \times 10^{-6}$~erg~cm$^{-2}$.
The equivalent burst duration, which is defined
as the ratio of burst fluence to peak flux, is estimated to be $\tau = 37.7 \pm 3.1$~s, consistent with the H-rich bursts previously observed from \target in the hard state~(see, e.g., Figure 5 in \citet{Takeda2025PASJ...77L..24T} for comparison with He-enhanced and pure He bursts).

For bursts with a complete profile, we estimated the individual burst fluences.
Each time-integrated spectrum over 0--150~s from the burst onset was fitted using the same spectral model described above, and the corresponding burst fluences are summarized in Table \ref{tab:burst_prop}.
The burst fluence in the soft state (ID 19) is roughly half of that in the hard state (IDs 1--14).

% -----------------------
\section{Discussion}
% -----------------------

\begin{figure*}[ht]
    % \vspace{-15pt}
    \begin{center}
        \includegraphics[width=170mm]{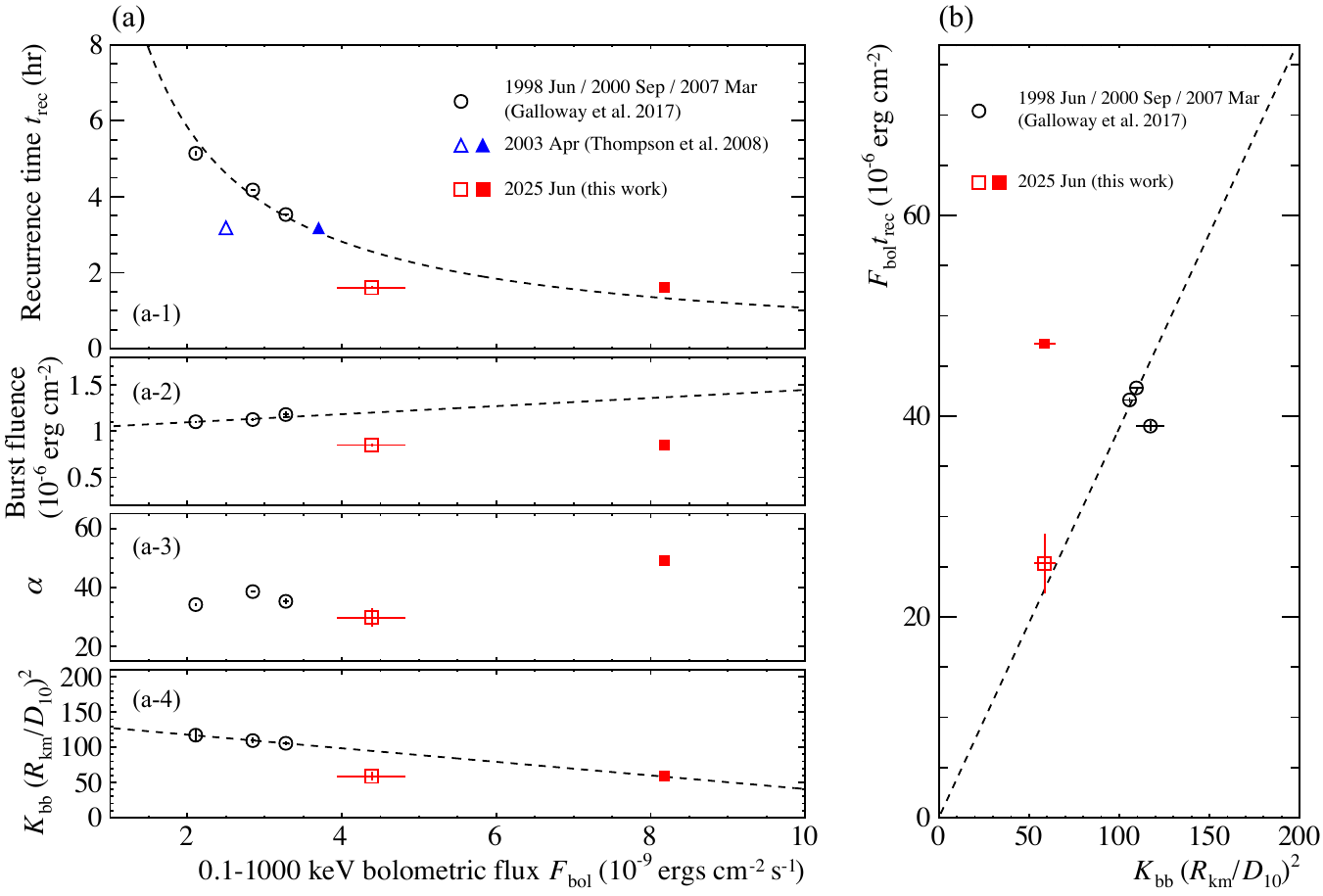}
        % \includegraphics[width=170mm]{figs/burst_Paramters_wSC3.pdf}
        % \includegraphics[width=180mm]{figs/burst_Paramters_wSC2.pdf}
        % \includegraphics[width=180mm]{figs/burst_Paramters_wSC.pdf}
        % \includegraphics[width=180mm]
        % \includegraphics[width=180mm]{figs/burst_Paramters.pdf}
        % % \includegraphics[width=180mm]{figs/burstParamters_vs_bolometricflux_gain1.059_3.pdf}
    \end{center}
    \caption{
    (a) Variations of (a-1) burst recurrence time, (a-2) burst fluence, (a-3) $\alpha$-value, and (a-4) blackbody normalization $K_{\rm bb}$ as a function of the bolometric persistent flux in the 0.1--1000~keV range.
    The dotted curve in panel (a-1) represents an empirical fit to the data from \citet{Galloway2017} (open black circles), assuming $t_{\rm rec} \propto F_{\rm per}^{-1.05}$ \citep{Galloway2004}.
    Panels (a-2) and (a-4) include linear fits to the same subset of data.
    The blue triangles represent the data from the 2003 April epoch reported by \citet{Thompson2008ApJ...681..506T}, while the red squares show the results from the 2025 June epoch. 
    In both cases, open and filled markers correspond to spectral fits without and with an additional disk blackbody component below 2~keV, respectively (see Section~\ref{sec:persistent} for details).
    % In both cases, open and filled markers correspond to spectral fits without and with an additional soft component below 2~keV, respectively (see Section~\ref{sec:persistent} for details).
    For the three previous epochs, $K_{\rm bb}$ values are the mean and standard deviation over 20--50~s after each burst onset, based on Figure~\ref{fig:BBcomponents_vs_time}(c). 
    For the 2025 June epoch, $K_{\rm bb}$ is derived from a spectral fit over the same interval.
    (b) 
    $F_{\rm bol}t_{\rm rec}$ plotted against $K_{\rm bb}$ during the burst tail.
    The black dotted line represents the best-fit linear model ($F_{\rm bol}t_{\rm rec} = a K_{\rm bb}$) to all data points excluding the 2025 June epoch results with additional soft component, yielding $a = 0.388 \pm 0.009$~$(R_{\rm km}/D_{10})^{-2}$erg~cm$^{-2}$ and $\chi^2/{\rm d.o.f} = 5.01/3$.
    } \label{fig:burstParamters_vs_bolometricflux_gain1.059}
\end{figure*}

\subsection{Exceptionally Short Burst Recurrence Time}
\label{sec:short_burst}
% \subsection{Historically Short Burst Recurrence Time}
% \subsection{Shortest Burst Recurrence Time}
% 観測結果のまとめと、過去との比較、$t_{\rm rec}$, $E_b$, $\alpha$、Table 2 (% Galloway+17 も含む表を作っても良いかも) にまとめる
% % α, mdot, yign, Xbar, 

During the NinjaSat observations, we discovered the shortest burst recurrence time in \target to date, $t_{\rm rec} = 1.603 \pm 0.040$~hr (Section~\ref{sec:burst}).
Theoretical models predict that the burst recurrence time should monotonically decrease with increasing accretion rate until stable burning sets in. 
However, observations of several bursters, including \target, reveal a “turn-over” regime, where the recurrence time increases again at higher accretion rates before bursting ceases~(e.g., \citealt{Cornelisse2003A&A...405.1033C, Galloway2008, Galloway2018ApJ...857L..24G, Cavecchi2020MNRAS.499.2148C}).
\citet{Cavecchi2025arXiv250611966C} interpret this behavior as a consequence of changes in the accretion flow geometry.
In \target, the previously reported burst recurrence time around the turn-over point was approximately 3~hr (see Figure 16 in \citealt{Galloway2008}).
Thus, the observed recurrence time of 1.6~hr is exceptionally short, indicating {\it anomalous} ignition conditions in the 2025 June epoch compared to previous epochs\footnote{Note that the shortest stable burst recurrence time to date, approximately 200~s, was observed in the 11 Hz X-ray pulsar IGR J17480$-$2446~\citep{Linares2012}.}.
This could provide a valuable observational constraint for theoretical models.

\subsection{Deviation from the Previous $t_{\rm rec}$--$F_{\rm bol}$ Relation}
Figure~\ref{fig:burstParamters_vs_bolometricflux_gain1.059}(a-1) shows the relation between the recurrence time and the bolometric persistent flux $F_{\rm bol}$ of \target, together with values from the previous three locked bursting epochs (1998 June, 2000 September, and 2007 March; \citealt{Galloway2017}), as well as from the 2003 April observations~\citep{Thompson2008ApJ...681..506T}.
In the former three epochs, the $t_{\rm rec}$--$F_{\rm bol}$ relation is well described by the power-law function (Equation~(\ref{eq:eta})), with $\eta = 1.05$~\citep{Galloway2004}.
The data point in 2003 April also aligns with this relation, considering the soft component below 2~keV in the spectral fit~(Section~\ref{sec:Introduction}; \citealt{Thompson2008ApJ...681..506T}).
Similarly, if we assume the presence of a soft component during the 2025 June epoch, our results broadly align with the previous relation, though with the caveat that NinjaSat is not fully sensitive to such low-temperature components due to its energy coverage above 2~keV~(Section~\ref{sec:persistent}).

On the other hand, if the soft component is not considered, the recurrence time we measured deviates significantly from the empirical relation: bursts occurred approximately 37\% earlier ($t_{\rm rec} = 1.603 \pm 0.040$~hr) than predicted from the persistent flux ($t_{\rm rec}= 2.55_{-0.24}^{+0.31}$~hr, based on the earlier relation). 
Furthermore, several clear differences remain when compared with the April 2003 epoch.
Whereas \citet{Thompson2008ApJ...681..506T} reported that burst fluences were comparable across previous epochs and that the $\alpha$-value\footnote{Ratio of accretion to thermonuclear energy release:\\ $\alpha = t_{\rm rec}F_{\rm bol}/E_{\rm b} = t_{\rm rec}F_{\rm per} c_{\rm bol}/E_{\rm b}, \label{eq:alpha} $ where $c_{\rm bol}$ is the bolometric correction factor, defined as the ratio of the bolometric persistent flux to that measured in the instrumental band.} was distinctly smaller in the 2003 April epoch, our 2025 June observations show a fluence reduced by $\approx 29\%$ relative to the linear trend established across previous epochs~(Figure~\ref{fig:burstParamters_vs_bolometricflux_gain1.059}(a-2)).
By contrast, the $\alpha$-value remains broadly consistent with earlier epochs (Figure~\ref{fig:burstParamters_vs_bolometricflux_gain1.059}(a-3)).

The reduction in the burst fluence can be attributed to the observed smaller blackbody normalization during the burst, which decreases by $\approx 38\%$ relative to the linear trend from earlier epochs (Figure\ref{fig:burstParamters_vs_bolometricflux_gain1.059}(a-4)).
Thus, unlike the 2003 April epoch, where the soft-component scenario could account for the observations, an alternative explanation is required for the 2025 June epoch to explain the above differences. 
This conclusion does not depend on whether a soft component below 2~keV is present, since the burst fluence is significantly lower than in previous epochs~(Figure\ref{fig:burstParamters_vs_bolometricflux_gain1.059}(a-2)).
If a soft component below 2~keV is indeed present, one possible interpretation of the observed differences is steady burning of part of the accreted material near the equator, with ignition triggered by unstable burning at higher latitudes (e.g., \citealt{Cavecchi2025arXiv250611966C}).
A summary of the observed burst properties of \target across four epochs is provided in Table~\ref{tab:burst_GS1826}.

\subsection{Physical Origin of the Short Recurrence Time}
% \subsection{Possible Origins of the Anomalously Short Recurrence Time}
\label{sec:short_burst_origin}

The fact that the 2025 June epoch exhibits {\it anomalously} short recurrence time (Section~\ref{sec:short_burst}) motivates us to explore possible physical origins, rather than attributing it solely to incomplete measurements of low-energy photons outside NinjaSat's coverage.
Here, we consider two scenarios, assuming no significant spectral component below 2~keV: additional heat flux from the NS crust, and a reduced accretion area on the NS surface.

An intriguing possibility is that the bursts ignited earlier due to an additional heat flux from the NS crust. 
Given that the crust was likely heated during the prolonged soft-state episode prior to the current observations~(Figure~\ref{fig:light_curves}(b, c)), and that its thermal timescale is on the order of a few months (e.g., \citealt{Brown2009ApJ...698.1020B,2017JPSJ...86l3901L}), the residual heat from the crust may be contributing to earlier ignition.
In time-dependent 1D burst simulation codes such as MESA~\citep{Paxton2015ApJS..220...15P} and KEPLER~\citep{Heger2007ApJ...671L.141H}, accretion-driven heating of the crust via non-equilibrium nuclear reactions, which is highly uncertain (e.g., \citealt{1990A&A...227..431H,2008A&A...480..459H,2018A&A...620A.105F}; see also \citealt{2021PhRvD.103j1301G} for the recent update), is parametrized as a base flux, $Q_{\rm b}$, or as a net base flux, $Q_{\rm e}$.
\citet{Zhen2023ApJ...950..110Z} show that recurrence time decreases by approximately 9\% for every 0.1~MeV~nucleon$^{-1}$ increase in $Q_{\rm e}$~(see their Figure 5 and Table 1){\footnote{We verified comparable dependence in multi-epoch modeling with KEPLER using the publicly available data set~\citep{Johnston2020_Mendeley}}.
The observed $\sim$40\% shorter recurrence time could be explained assuming an increase in $Q_{\rm e}$ of roughly 0.4~MeV~nucleon$^{-1}$ relative to the values in previous epochs.
Nevertheless, a comprehensive modeling of the crust's thermal evolution after the soft-to-hard transition, along with the subsequent clocked bursting phase, is crucial to test this hypothesis (Dohi et al. in prep.), although this scenario alone may not be sufficient to explain the observed reduction in the blackbody normalization.

One plausible explanation is that accretion confined to a smaller fraction of the NS surface leads to earlier ignition.
As pointed out by \citet{Bildsten2000AIPC..522..359B}, the burst recurrence time is determined by the local accretion rate per unit area, $\dot{m}$, rather than by the global accretion rate, $\dot{M}$.
Assuming a uniform local accretion rate over the accreted area, the total accreted mass until ignition, $\dot{M}t_{\rm rec}$, is proportional to the accreted area, $S$, and can be expressed as
\begin{equation}
    \dot{M}t_{\rm rec} = \dot{m}St_{\rm rec} = y_{\rm ign}S,
    \label{eq:Mass_K}
\end{equation}
where $y_{\rm ign} = \dot{m}t_{\rm rec}$ is a critical column depth for ignition.
If we assume that the blackbody normalization during the burst tail, which is proportional to the emitting area, is a good indicator of the accreted area ($S \propto K_{\rm bb}$) and that the bolometric persistent flux is proportional to the global accretion rate ($F_{\rm bol} \propto \dot{M}$), Equation~(\ref{eq:Mass_K}) can be rewritten as 
\begin{equation}
    F_{\rm bol}t_{\rm rec}  \propto K_{\rm bb}.
    \label{eq:Mass_K2}
\end{equation}
Assuming no significant spectral component below 2~keV, Equation~(\ref{eq:Mass_K2}) naturally accounts for the observed burst properties in \target, whereas it cannot explain the observed differences if the soft component is considered~(Figure~\ref{fig:burstParamters_vs_bolometricflux_gain1.059}(b)).
These results support a scenario in which the soft component is negligible in the 2025 June epoch, although its contribution cannot be fully ruled out.
The reduced blackbody normalization observed in our data suggests that both fuel accumulation and burning were confined to a narrower region on the NS surface.
This interpretation is further supported by the persistent spectral parameters, particularly the lower electron temperature ($kT_e = 6.1^{+1.7}_{-0.9}$~keV) and higher optical depth ($\tau = 4.72_{-0.6}^{+0.4}$ for disk geometry) compared to typical hard-state values~\citep{Cocchi2010A&A...509A...2C,Yun2023ApJ...947...81Y}. 
These parameters suggest that the corona became optically thicker and geometrically thinner just prior to the transition to the soft state, possibly restricting the accretion flow near the equator.

We note that the anti-correlation between the blackbody normalization and the persistent flux in \target\ was reported during three previous epochs~\citep{Galloway2004}, as also shown in Figure~\ref{fig:burstParamters_vs_bolometricflux_gain1.059}(a-4). \citet{Galloway2012ApJ...747...75G} suggested that this trend may arise from an anisotropy of the burst emission or variations in the spectral color correction factor, $f_c$.
However, we stress that the reduced accretion area scenario provides a more plausible explanation for the {\it anomalously} short burst recurrence time observed in our 2025 June sample, although the above effects may partially affect the measured blackbody temperature and normalization, not the recurrence time.
Our results underscore that even in \target, where observation–model comparisons have been most successful, more realistic modeling is beneficial, accounting for multi-dimensional accretion as well as crust thermal evolution.

% \clearpage

\appendix % 付録部分の始まり
\section{Properties of the persistent and burst emissions observed from \target}

\renewcommand{\thetable}{A\arabic{table}}
\setcounter{table}{0}  % 付録開始時に表番号をリセット

The persistent and burst emissions are characterized by several parameters, including the recurrence time ($t_{\rm rec}$), absorbed persistent flux in the 3--25~keV band ($F_{\rm per}$), bolometric correction factor ($c_{\rm bol}$), burst fluence ($E_{\rm b}$), peak flux ($F_{\rm pk}$), $\alpha$-value, and blackbody normalization during the burst tail ($K_{\rm bb}$).
Table~\ref{tab:burst_GS1826} lists these parameter values, with the current observational results added to those from three previous epochs (1998 June, 2000 September, and 2007 March) taken from \citet{Galloway2017}.
$c_{\rm bol}$ and the $\alpha$-value are estimated under the assumption that no soft component exists below 2~keV~(see Section\ref{sec:persistent}).

{
\tabcolsep = 3.1pt
\begin{table*}[h]
\centering
% \small
\footnotesize
% \scriptsize
\begin{threeparttable}

\caption{Comparison of Type-I X-ray burst properties from GS 1826$-$238 across four epochs.}
% \caption{Properties of Type-I X-ray bursts from GS 1826$-$238 observed in four epochs.}
\label{tab:burst_GS1826}
\begin{tabular*}{\textwidth}{@{\extracolsep{\fill}}lcccccccc}
% \begin{tabular}{cccccccc}
% \begin{tabular}{c*{8}{c}}
\hline
\hline
Epoch & 
$t_{\rm rec}$ (hr)&
{\parbox[c]{2.8cm}{\centering $F_{\rm per}$ \\ (10$^{-9}$~erg cm$^{-2}$ s$^{-1}$)}} &
$c_{\rm bol}$ &
{\parbox[c]{2.6cm}{\centering $E_{\rm b}$ \\ (10$^{-6}$~erg cm$^{-2}$)}}&
{\parbox[c]{2.8cm}{\centering $F_{\rm pk}$ \\ (10$^{-9}$~erg cm$^{-2}$ s$^{-1}$)}}&
$\alpha$ &
{\parbox[c]{2cm}{\centering $K_{\rm bb}$ \\ $(R_{\rm km}/D_{10})^{2}$ }}& \\
\hline
1998 Jun & 5.14 $\pm$ 0.07 & 1.167 $\pm$ 0.006 & 1.806 $\pm$ 0.009 & 1.102 $\pm$ 0.011 & 30.9 $\pm$ 1.034 & 34.2 $\pm$ 0.5 & $117.2 \pm 8.0$\\
2000 Sep & 4.177 $\pm$ 0.010 & 1.593 $\pm$ 0.017 & 1.787 $\pm$ 0.003 & 1.126 $\pm$ 0.016 & 29.1 $\pm$ 0.538 & 38.6 $\pm$ 0.3 & $109.5 \pm 4.0$\\
2007 Mar & 3.530 $\pm$ 0.004 & 1.87 $\pm$ 0.02 &  1.751 $\pm$ 0.003 & 1.18 $\pm$ 0.04 & 28.4 $\pm$ 0.4 & 35.3 $\pm$ 1.0 & $105.7 \pm 3.1$\\
\midrule
{\parbox[c]{1.5cm}{2025 Jun  \\ (this work)}} 
& $1.603 \pm 0.040$ & $2.707_{-0.083}^{+0.056}$ & $1.62 \pm 0.17$ & $0.849 \pm 0.019$ & $22.5 \pm 1.8$ & $29.8 \pm 3.2$ & $58.6_{-6.0}^{+6.7}$\\
\hline
\end{tabular*}
% \begin{tablenotes}
% \item \textbf{Note.} 
% The values for the three previous epochs are taken from \citet{Galloway2017}. 
% % $c_{\rm bol}$ is estimated based on the absorbed 3–25 keV and unabsorbed bolometric 0.1–1000 keV fluxes in the hard state.
% \end{tablenotes}
\end{threeparttable}
\end{table*}
}

\begin{acknowledgments}
This project was supported by JSPS KAKENHI (JP17K18776, JP18H04584, JP20H04743, JP20H05648, JP21H01087, JP23K19056, JP24H00008, JP24K00673, JP25H01273, JP25KJ0241, JP25K17403). 
T. Takeda acknowledges support from the JSPS Research Fellowships for Young Scientists (JSPS KAKENHI Grant Number JP25KJ0241).
T.E. was supported by the “Extreme Natural Phenomena” RIKEN Hakubi project, and the JST Japan grant number JPMJFR202O (Sohatsu).

\end{acknowledgments}

\vspace{5mm}
\facilities{NinjaSat, MAXI}

\bibliography{reference}{}
\bibliographystyle{aasjournal}

% ===========

\end{document}